\documentclass[12pt]{article}
\usepackage{fullpage,setspace}
\usepackage{amssymb, amsthm, amsmath}
\usepackage{graphicx}
\usepackage[authoryear]{natbib}
\usepackage{bm}
\usepackage{lineno}
\usepackage{comment}
\usepackage{caption}
\usepackage{subcaption}
\usepackage{mathtools}
\usepackage{multirow}
\usepackage{gensymb}
\usepackage[title]{appendix}

\usepackage[T1]{fontenc}
\usepackage[sc]{mathpazo}

\usepackage{algorithm}
\usepackage{algorithmicx}
\usepackage[ruled,vlined,algo2e]{algorithm2e}
\usepackage{hyperref}
\usepackage{booktabs}
 \usepackage{relsize}

\newcommand{\beq}{ \begin{equation}}
\newcommand{\eeq}{ \end{equation}}
\newcommand{\beqn}{ \begin{eqnarray}}
\newcommand{\eeqn}{ \end{eqnarray}}

\usepackage{caption}
\begin{document} 
\newtheorem{res}{Result}
\newcommand{\myvec}[1]{\pmb{#1}}
\newcommand{\deri}[1]{\frac{\partial}{\partial #1 }}
\newcommand\norm[1]{\left\lVert#1\right\rVert}
\newcommand{\pkg}[1]{{\fontseries{b}\selectfont #1}} 
\newcommand{\distas}[1]{\mathbin{\overset{#1}{\kern\z@\sim}}}%
\newsavebox{\mybox}\newsavebox{\mysim}
\newcommand{\distras}[1]{%
  \savebox{\mybox}{\hbox{\kern3pt$\scriptstyle#1$\kern3pt}}%
  \savebox{\mysim}{\hbox{$\sim$}}%
  \mathbin{\overset{#1}{\kern\z@\resizebox{\wd\mybox}{\ht\mysim}{$\sim$}}}%
}

\begin{center}
{\Large Optimal Stock Portfolio Selection with a Multivariate Hidden Markov Model}\\\vspace{6pt}
{\large Reetam Majumder\footnote[1]{University of Maryland, Baltimore County}, Qing Ji\footnote[2]{Procter \& Gamble} and Nagaraj K. Neerchal$^1$}\\
\end{center}

\begin{abstract}\begin{singlespace}\noindent
The underlying market trends that drive stock price fluctuations are often referred to in terms of bull and bear markets. 
Optimal stock portfolio selection methods need to take into account these market trends; however, the bull and bear market states tend to be unobserved and can only be assigned retrospectively. We fit a linked hidden Markov model (LHMM) to relative stock price changes for S\&P 500 stocks from 2011--2016 based on weekly closing values. The LHMM consists of a multivariate state process whose individual components correspond to HMMs for each of the 12 sectors of the S\&P 500 stocks. The state processes are linked using a Gaussian copula so that the states of the component chains are correlated at any given time point. The LHMM allows us to capture more heterogeneity in the underlying market dynamics for each sector.
In this study, stock performances are evaluated in terms of capital gains using the LHMM by utilizing historical stock price data.
Based on the fitted LHMM, optimal stock portfolios are constructed to maximize capital gain while balancing reward and risk. Under out-of-sample testing, the annual capital gain for the portfolios for 2016--2017 are calculated. Portfolios constructed using the LHMM are able to generate returns comparable to the S\&P 500 index.

\vspace{12pt}
\end{singlespace}
\end{abstract}

\section{Introduction}\label{intro}
A stock portfolio refers to a collection of stocks selected and owned by an investor, and
stock portfolio selection has been at the center of investment methodology research for many years. Depending on the investment goals, various methods have been developed by researchers for selecting stocks and allocating assets. The modern portfolio selection methodology developed by \citet{M52} has guided a large section of portfolio research. There are two essential components to the portfolio selection procedure, namely the evaluation of stocks, and portfolio assets allocation. A good introductory reference for topic of portfolio selection is \citet{M19}. In this paper, we follow the groundwork laid out by \citet{JN19} of connecting portfolio selection to the estimation of the underlying statistical model. We first build statistical models using past data on stock prices. Then, optimal stock portfolios are constructed based on the technique in \citet{M52} to maximize the capital gain while balancing reward and risk. The performance of the optimal portfolios are evaluated by comparing annual gains based on the portfolios against the S\&P 500 gains for the same time period.

    Stock markets around the world use the terms bull and bear to describe market trends. Stock prices are relatively stable and generally increasing in a bull market. A bear market, on the other hand, indicates strong market volatility with decreasing stock prices.  A bull to bear market switch or vice versa is recognized after an increase/decrease of 20\% or more in multiple stock indices \citep{KD16}. While bull and bear markets cannot be directly observed, the behaviour of individual stocks point to the state of the market. The current state of a stock can be estimated by analysts, but the true state is unknown unless evaluating stocks retrospectively. Therefore, the state of a stock can be treated as an unobserved (latent) random variable, and the prices of the stock are the observed values. In addition, the market conditions can switch states at any time point. Given these characteristics, a hidden Markov model (HMM) is well suited for modeling the bull/bear trend of the market.  

\begin{figure}
  \centering
    \includegraphics[width=.7\linewidth]{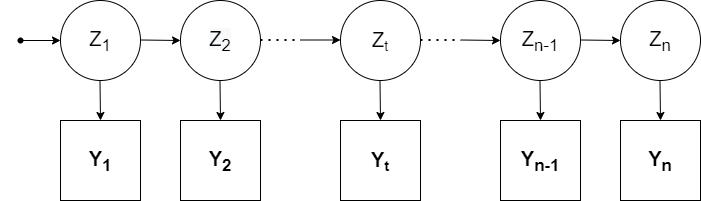}
    \caption{\small A directed acyclic graph (DAG) specifying the conditional independence structure for a hidden Markov model.}\label{fig:HMM}
\end{figure}
An HMM is a discrete-time stochastic process that is controlled through a Markov chain with latent (hidden) states.  
A Markov chain (MC) is a well-known stochastic model that describes a sequence of discrete events.  Let a sequence of random variables $(Z_1,Z_2, \ldots, Z_n)$ form a Markov chain.  The characterizing property of a first order Markov chains states that 
\begin{equation}\label{Memoryless}
P(Z_t=z_t \mid Z_{t-1}=z_{t-1}, \ldots, Z_1=z_1) = P(Z_t=z_t  \mid  Z_{t-1}=z_{t-1}).  
\end{equation}
Assuming the Markov chain is stationary and has $J$ states, the transition probabilities $P(Z_t=z_t  \mid  Z_{t-1}=z_{t-1})$ can be arranged into a $J$ by $J$ matrix known as the transition probability matrix,
	\[{\bf{\Pi}}=
  \left[ {\begin{array}{cccc}
   \pi_{11} & \pi_{12} & \cdots & \pi_{1J}\\
   \pi_{21} & \pi_{22} & \cdots & \pi_{2J}\\
   \vdots & \vdots & \ddots & \vdots \\
   \pi_{J1} & \pi_{J2} & \cdots & \pi_{JJ}\\
  \end{array} } \right],
\]
where $\pi_{hj}=P(Z_t=j \mid Z_{t-1}=h)$ is the probability of transitioning from state $h$ to state $j$ at any $t$, and $\sum\limits^J_{j=1} \pi_{hj} = 1$ given any $h$.  In an HMM, observations are assumed to be drawn from one of several sub-distributions determined by the unobserved variable $Z_t$.  Formally, an HMM consists of a pair of random processes $\{Z_t,Y_t\}_{t\geq 1}$, where $\{Z_t\}$ is a Markov chain with $J$ states.  Conditional on $\{Z_t\}$, $\{Y_t\}$ is a sequence of independent random variables such that the distribution of $Y_t$ depends only on $Z_t$. The conditional distribution of $Y_t$ given $Z_t$ is given by
\begin{equation}\label{HMM_EX}
Y_t \mid Z_t=j \stackrel{}{\sim}  f_j (y \mid\myvec{\theta}_j),\mbox{\hspace{5pt}}j=1,\ldots, J,
\end{equation}
where $f_1,\ldots,f_J$ are the different sub-distributions.  $\{Z_t\}$ is known as the state process of the HMM, and $\{Y_t\}$ is known as the emission process. Note that at any time point $t$, $Y_t$ could be distributed as a univariate or as a multivariate distribution. Figure \ref{fig:HMM} depicts the graphical representation of an HMM. The variables in this representation are denoted as the nodes of the graph, and the arrows connecting them are denoted as the edges and represent the dependence among the nodes.

Some previous work on predicting stock prices based on HMMs include \citet{BH05}, and \citet{N18}.  Their models were trained directly using the stock closing values and were used to predict stock prices in the near future; there is however no extension to portfolio selection in their work. \citet{H89} combined the HMM structure with autoregressive models in order to capture the market trend, where parameters of an autoregressive model were considered to arise from an HMM.  \citet{EH97} and \citet{ESB10} further extended the work of \citet{H89} to include a portfolio selection procedure.
\begin{figure}
  \centering
    \includegraphics[width=.4\linewidth]{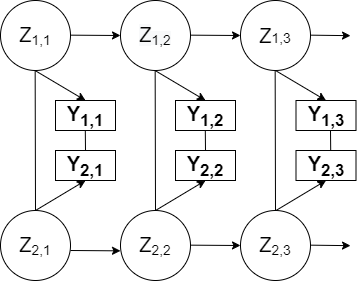}
    \caption{\small Graphical representation of 3 time slices of a linked hidden Markov model (LHMM) with 2 state processes specified as a multivariate Markov chain (MMC), where each Markov chain affects only a partition of the emission process.}\label{fig:LHMM4}
\end{figure}
Although we may largely expect the bull and bear states of the market to be consistently reflected in weekly stock price changes, it is likely that stocks in different sectors will have different underlying dynamics of the state process. This paper considers a model where each sector is driven by a different state process with its own bull and bear states. This results in a multivariate state process, whose individual components are Markov chains. The dependence between the individual Markov chains of the multivariate Markov chain (MMC) can propagate in different ways, and \citet{MajumderThesis} discusses some of the common ways an MMC has been specified in previous studies. Additonal work in the area of HMMs and correlations among the prices of different assests or markets include \citet{EK14}, who investigated the correlations between different stock sectors while applying a regime-switch model to the correlation matrix, and \citet{FFST14}, who address the estimation of a multivariate HMM using shrinkage estimators. More recently, \citet{XC21} have incorporated a vine-coupla into an artificial neural network so that the inter-market correlations were considered while estimating the return of a portfolio. We thank one of the anonymous referees for bringing these references to our attention.

In this paper, we assume that the state processes of the MMC evolve in lockstep, i.e., the nodes of two Markov chains are connected by an edge if and only if the nodes are at the same time point. The resulting multivariate HMM is known as a linked HMM (LHMM). Figure \ref{fig:LHMM4} represents our approach through an example where $\textbf{Z}_t' = (Z_{1,t},Z_{2,t})$ is a bivariate state process corresponding to bull/bear states for 2 sectors of the stock market, and $\textbf{Y}_t' = (\textbf{Y}_{1,t},\textbf{Y}_{2,t})$ are the stock returns for all stocks within the two sectors. This is a modification of the default LHMM specification; the state processes of the different sectors can be considered to evolve in lockstep, and each state process affects the stock price changes for that sector's stocks. Partitioning the stocks by sector allows for more heterogeneity in the market dynamics while still using two-state latent processes with an intuitive bull/bear labeling. We can extend this idea to an LHMM with $D$ clusters corresponding to $D$ sectors. We specify the dependency structure for the $D$-variate LHMM using a Gaussian copula, which allows us to generate correlated states from the MMC at every time point \citep{MajumderThesis}. To demonstrate our LHMM, we propose a stock portfolio selection method based on the work of \citet{JN19}.  

The rest of this paper is structured as follows. In Section~2, we describe an LHMM with its dependency structure specified using a Gaussian copula which can be used to model weekly stock price changes. Section~3 introduces the methods to evaluate stocks and portfolios and explains the portfolio selection methodology. In Section~4, we validate the portfolio selection method using historical S\&P 500 stock data. Finally, Section~5 discusses our results and proposes ways that our approach can be improved.


\section{Parameter Estimation for a Linked Hidden Markov Model}\label{sec:param-estimation}
\subsection{Parameterizing an LHMM using a Gaussian copula}\label{sec:copula}
Suppose a stock portfolio consists of $K$ stocks. 
For the $k$\textsuperscript{th} stock, $k=1,2,\ldots,K$, let $X_{k,t}$ be the closing price at the end of the $t$\textsuperscript{th} week, $t=1,2,
\ldots,n$.  The price changes in percentage are given by 
\begin{equation*}
Y_{k,t}=\frac{X_{k,t}-X_{k,t-1}}{X_{k,t-1}}.  
\end{equation*}
Let $\textbf{Y}_t = (Y_{1,t},\ldots,Y_{K,t})$ be the vector of stock price changes at the end of the $t$\textsuperscript{th} week, and $Z_{t}$ be the binary latent state at that time point. The HMM for the stock price changes is given by
\begin{equation}\label{HMM}
 \textbf{Y}_{t} \mid Z_{t}=j \stackrel{}{\sim} \prod_{k=1}^K N(\mu_{k,j} , \sigma^2_{k,j}),\mbox{\hspace{5pt}}j = 1,2,
\end{equation}
where $(Z_{1},\ldots,Z_{n})$ is a Markov chain with initial distribution $\myvec{\alpha}$ and a $2\times2$ transition matrix ${\bf{\Pi}}$. The latent states represent the bull or bear state of the market, which is observed in the emission process as the buy or sell trend for each of the stocks. The emission distribution assumes a conditional independence structure at each time point, i.e., the price changes of any stock is independent of the remaining stocks conditional on the state. Following notation established in \eqref{HMM_EX}, we use $\myvec{\theta}$ to denote all the parameters of the emission distribution. Parameter estimation for an HMM of this form is carried out using the Baum-Welch (B-W) algorithm \citep{BP66}, which is a special case of the expectation-maximization (EM) algorithm \citep{DLR77}. A comprehensive tutorial of parameter estimation in HMMs is provided by \citet{Rabiner89}.

Now, let us consider the case where the $K$ stocks belong to $D$ different sectors. If we assign each sector its own underlying state process, we can denote the state of LHMM at the end of the $t$\textsuperscript{th} week as $\textbf{Z}_t = (Z_{1,t},\ldots, Z_{D,t})$. If the $d$\textsuperscript{th} sector consists of $n_d$ stocks, the HMM for price changes in the $d$\textsuperscript{th} sector is given by,
\begin{align}\label{eqn:lhmmEmissions}
    \mathbf{Y}_{d,t}\vert Z_{d,t} = j \sim \prod_{k_d=1}^{n_d} N(\mu_{k_d,j},\sigma^2_{k_d,j}),\mbox{\hspace{5pt}} j=1,2,
\end{align}
with $\sum_{d=1}^D n_d=K$.
As before, $Z_d'=(Z_{d,1}, \ldots, Z_{d,n})$ is the latent state process. The HMM for sector $d$ is parameterized by the initial distribution $\myvec{\alpha}_d$, transition matrix $\mathbf{\Pi}_d$, and emission distribution parameters $\myvec{\theta}_d$. Furthermore, $\{Z_1,\ldots,Z_D\}$ is a $D$-component MMC, and $Y_{d,t}$ given $Z_{d,t}$ is independent to $Y_{d',t'}$ and $Z_{d',t'}$ for any $d' \neq d$. The full likelihood of the LHMM at time \textit{t} can be written as,
\begin{align}\label{eqn:lik1}
    f(\mathbf{Y}_{1,t},\ldots,\mathbf{Y}_{D,t}, Z_{1,t},\ldots, Z_{D,t}) = f(Z_{1,t},\ldots, Z_{D,t}) \prod_{d=1}^Df(\mathbf{Y}_{d,t}\vert Z_{d,t}),
\end{align}
where the parameter dependencies have been suppressed for convenience. We want to parametererize the association between the component Markov chains of the MMC at every time point, and our approach to that end is to construct a Gaussian copula for the state processes. Let $F$ be the \textit{D}-dimensional joint CDF of $\{Z_1,\ldots,Z_D\}$, and let $F_1, \ldots, F_D$ be the marginal CDFs of $Z_1, \ldots, Z_D$ respectively. We define a Gaussian copula over the state processes as:
\begin{align}
    F(Z_1, \ldots, Z_D) &= \mathcal{C}\bigl(F_1(Z_1;\myvec{\alpha}_1,\mathbf\Pi_1), \ldots, F_D(Z_D;\myvec{\alpha}_D,\mathbf \Pi_D)\bigr)\nonumber \\
                               &= \mathlarger{\Phi}_D\bigl(\Phi^{-1}(U_1), \ldots, \Phi^{-1}(U_D);\Sigma\bigr) \nonumber\\
                               &= \mathlarger{\Phi}_D\bigl(W_1, \ldots, W_D;\Sigma\bigr) \label{eqn:lhmmcopula},
\end{align}
where $U_1,\ldots, U_D$ are $Uniform(0,1)$ variates and $W_1, \ldots W_D$ are standard Normal variates. $\mathlarger{\Phi}_D$ is a $D$-dimensional multivariate Normal CDF with correlation matrix $\Sigma$, while $\Phi^{-1}$ is the inverse CDF of a univariate standard Normal distribution. Note that $W_d = \Phi^{-1}\bigl(F_d(Z_d;\myvec{\alpha}_d,\mathbf \Pi_d)\bigr)$, and therefore the joint distribution of the state processes can be obtained by using the chain rule as,
\begin{align}
    f(Z_1,\ldots,Z_D) &= \frac{\mathlarger{\phi}_D\bigl(W_1,\ldots,W_D;\Sigma\bigr)}{\phi(W_1)\times \ldots \times \phi(W_D)}\prod_{d=1}^D f_d(Z_d) \nonumber \\
    &= c(Z_1,\ldots,Z_D;\Sigma)\prod_{d=1}^D f_d(Z_d) \label{eqn:copulapdf},
\end{align}
where $\phi_D$ and $\phi$ are the density functions corresponding to $\Phi_D$ and $\Phi$, $f_d(\cdot)$ denotes the distribution of $Z_d$, and $c(\cdot)$ denotes the copula density. The likelihood in \eqref{eqn:lik1} can thus be simplified to
\begin{align}
    f(\mathbf{Y}_{1,t},\ldots,\mathbf{Y}_{D,t}, Z_{1,t},\ldots, Z_{D,t}) &= c(Z_{1,t},\ldots, Z_{D,t})
    \prod_{d=1}^Df(\mathbf{Y}_{d,t}\vert Z_{d,t})f_d(Z_{d,t})\nonumber \\
    &= c(Z_{1,t},\ldots, Z_{D,t})
    \prod_{d=1}^Df(\mathbf{Y}_{d,t}, Z_{d,t}).
\end{align}
The copula augmented model has an LHMM structure similar to Figure \ref{fig:LHMM4}. A discussion of the fundamentals and theoretical properties of copulas can be found in \citet{N06}. Copulas of continuous variables are well defined and have been extensively studied, but constructing a copula for discrete variables is not as straightforward. Since a Markov chain is either a nominal or an ordinal random variable, finding an appropriate measure of association between latent state processes to construct a copula can be challenging. To address this, we will take advantage of a unique relationship which exists between the Spearman and Pearson correlations of a bivariate Normal distribution. Since the Spearman correlation can be used as a measure of association for ordinal data, we choose a Gaussian copula parameterized by a correlation matrix $\Sigma$. We assume that the state processes evolve in lockstep, and correlated $D$-vectors from  $\Phi_D(\cdot \mid \Sigma)$ can be linked to an MMC with correlated Markov chains by means of an appropriate transformation. One such method is described below.

\subsection{Constructing an MMC from Uniform random variates}\label{sec:serfozoji}
Since the copula is a $D$-dimensional CDF with Uniform marginals, we discuss a method to generate an MMC from a Gaussian copula in this section. This will be relevant to our method of estimating the copula parameters.

Let us first review a method to generate a univariate Markov chain; \citet{S09} describes how to construct a Markov chain from a Uniform variable.
Suppose that the desired Markov chain has the initial distribution vector $\myvec{\alpha}$ and the transition probability matrix ${\bf \Pi}$.  Let $h(u)$ and $f(j,u)$ be functions transforming continuous values into categorical values $\mathcal{J}=\{1,2,\ldots,J\}$.  They are given by
\begin{equation}\label{h_f}
h(u) = j \text{   if } u\in I_j\text{ for some }j\in\mathcal{J},
\end{equation}
where $I_1=\left[0,\alpha_1\right)$ and $I_j=\left[\sum^{j-1}_{l=1}\alpha_l,\sum^{j}_{l=1}\alpha_l\right)$ for any $j>1$, and
\begin{equation}\label{f_f}
f(i,u) = j \text{   if } u\in I_{ij}\text{ for some }j\in\mathcal{J},
\end{equation}
where $I_{i1}=\left[0,\pi_{i1}\right)$ and $I_{i,j}=\left[\sum^{j-1}_{l=1}\pi_{il},\sum^{j}_{l=1}\pi_{il}\right)$ for any $j>1$.

Let ${\bf U}=(U_1,U_2,\ldots,U_n)$ be a vector of independent random variables where $U_t$ has an uniform distribution on $[0,1]$.  We will denote $h(U_1)$ as $Z_1$ and $f(Z_{t-1},U_{t})$ as $Z_{t}$ for any $t>1$. \citet{S09} showed that $(Z_1,Z_2,\ldots,Z_n)$ is a Markov chain with the initial distribution $\myvec{\alpha}$ and the transition probability matrix ${\bf \Pi}$.  \citet{JiThesis2019} modified this method in order to generate correlated Markov chains.
In a univariate Markov chain, the random value at $t$\textsuperscript{th} time point, $Z_t$, is generated from a single random variable $U_t$.
To create an MMC with $D$ Markov chains, we need a vector of possibly correlated random variables $(U_{1,t},\ldots,U_{D,t})$ at each time point $t$.  Therefore, we will use a $D$-dimensional Normal distribution to generate correlated random values. 
Suppose that an MMC has a length of $n$ with $D$ sequences.   Let the stationary distribution and the transition probability matrix of the $d$\textsuperscript{th} sequence be $\myvec{\eta}_d$ and ${\bf \Pi}_d$ respectively.  We use the inverse transform method to create Uniform variables from Normal variables \citep{R19}.
For the $t$\textsuperscript{th} time step, $1 \leq t \leq n$, let ${\bf W}_t = (W_{1,t},\ldots,W_{D,t})$ and ${\bf W}_t \overset{\text{i.i.d}}{\sim} \text{MVN}(\myvec{0},{\Sigma})$ where ${\Sigma}$ is a correlation matrix.  
For each $d = 1,\ldots,D$ and $t = 1,\ldots,n$, a Uniform random variable is created using the inverse transform method, namely $U_{d,t} = \Phi(W_{d,t})$. Each $U_{d,t}$ is thus uniformly distributed on $[0,1]$.  
The correlations among $U_{1,t},\ldots,U_{D,t}$ stem from the correlations among $W_{1,t},\ldots,W_{D,t}$.  

Now let us apply (\ref{h_f}) and (\ref{f_f}) to $U_{d,t}$.  A random variable $Z_{d,t}$ is created for each $(d,t)$ pair where $Z_{d,1}=h(U_{d,1})$ and $Z_{d,t}=f(Z_{d,t-1},U_{d,t})$. Thus, we have an MMC $\{{\bf Z}_1, \ldots, {\bf Z}_n\}$ where ${\bf Z}_t=(Z_{1,t},Z_{2,t},\ldots,Z_{D,t})$ and $\{ Z_{d,1},Z_{d,2},\ldots,Z_{d,n} \}$ is a Markov chain marginally.  In addition, $Z_{1,t},\ldots, Z_{D,t}$ are correlated at the $t$\textsuperscript{th} time step. The functions in (\ref{h_f}) and (\ref{f_f}) are collectively referred to as $g(\cdot)$ going forward, and describes the overall process of transforming marginally Uniform random vectors into an MMC.

\subsection{Two-stage parameter estimation for the LHMM}\label{sec:paramestimation}
The construction of a copula for the state processes requires knowledge of the states that give rise to the data. This is usually obtained as the most likely sequence of states using the Viterbi Algorithm \citep{Viterbi}. The Viterbi Algorithm is applied after the model parameters have been estimated - this means that we need to resort to a two-stage estimation process. In the first stage, the parameters for each sector's HMMs are estimated independently using the B-W algorithm. The Viterbi Algorithm then provides us the most likely sequence of states to have generated the data, which is used to estimate the copula correlation matrix $\Sigma$. Afterwards, the marginal parameters can be re-estimated conditioned on the correlation structure.

Estimating $\Sigma$ in \eqref{eqn:lhmmcopula} is challenging using conventional approaches like the inversion method \citep{N06} or the inference functions for margins method \citep{Joe1996TheEM}, since neither the CDF $F_d(Z_d)$ nor its associated probability mass function that appear in \eqref{eqn:lhmmcopula} and \eqref{eqn:copulapdf} can be evaluated easily. Instead, we choose $\Sigma$ in a manner such that states generated from the Gaussian copula using the methodology discussed in Section \ref{sec:serfozoji} can be used to reproduce a desired measure of association for the MMC. For each stock, we assumed that the HMM has two hidden states, the bear state and the bull state. However, the B-W algorithm produces two states, State 1 and State 2 without labels identifying them as bear/bull. So without loss of generality, we relabel the states for the $d$\textsuperscript{th} Markov chain such that for State 1,
\begin{align*}
 \sum_{k_d=1}^{n_d}\dfrac{\mu_{k_d,1}}{\sigma_{k_d,1}} > \sum_{k_d=1}^{n_d}\dfrac{\mu_{k_d,2}}{\sigma_{k_d,2}},   
\end{align*}
where $\mu_{k_d,1}, \mu_{k_d,2},\sigma_{k_d,1},\mbox{ and }\sigma_{k_d,1}$ are as defined in \eqref{eqn:lhmmEmissions}. State 1 has a higher return to volatility ratio and can be considered a good stock to buy \citep{HH15}. It would thus correspond to a bull market, and State 2 can be considered to be bear market states. Since the states are now ordinal in nature, the pairwise Spearman correlation for Markov chains in the MMC is chosen as the desired measure of association. However, there is no obvious way to estimate a $D \times D$ matrix $\Sigma$ whose pairwise correlations are functions of the Spearman correlations between the Markov chains. We make a simplifying assumption for the copula and rewrite \eqref{eqn:lhmmcopula} as:
\begin{align}
    \mathlarger{\Phi}_D\bigl(W_1,\ldots, W_D; \Sigma \bigr) \approx \prod_{d_1=1}^{D-1} \prod_{d_2 = d_1+1}^D \mathlarger{\Phi}_2(W_{d_1},W_{d_2}; \rho_{d_1d_2}), \label{eqn:gaussianpaircopula}
\end{align}
where $\rho_{d_1d_2}$ denotes the Pearson correlation between $W_{d_1}$ and $W_{d_2}$, and corresponds to the $(d_1,d_2)$\textsuperscript{th} element of $\Sigma$. This formulation can be interpreted in a manner similar to a pairwise simplified regular vine (R-vine) copula \citep{Brechmann2012}, with all pair-copula terms involving a conditioning set replaced by bivariate Gaussian copulas. We refer to this as the pair-copula approximation, and it consists of $D(D-1)/2$ terms. The copula density associated with \eqref{eqn:gaussianpaircopula} can also be interpreted as a composite likelihood \citep{varin2011}. In practice, this will allow us to estimate the individual elements $\rho_{d_1d_2}$ of $\Sigma$ using the right hand side of \eqref{eqn:gaussianpaircopula}, but simulate data from the copula using the left hand side of \eqref{eqn:gaussianpaircopula}, as long as we can ensure that $\Sigma$ is a positive-definite matrix. \citet{kruskal1958} provided a relationship between the Pearson correlation $\rho$ and the Spearman correlation $\rho^*$ for bivariate Normal variables $(W_1,W_2)$ that we will use to estimate  $\rho_{d_1d_2}$:
\begin{equation}\label{eq:spearman}
    \rho = 2 \sin\biggl{[}\pi\frac{\rho^*}{6}\biggr{]}.
\end{equation}
Note that $\rho^*(W_{d_1},W_{d_2}) = \rho^*(U_{d_1},U_{d_2})$ since the Spearman correlation coefficient is invariant under monotone transforms. Recall that we defined $g(\cdot)$ in Section \ref{sec:serfozoji} as the function which transforms Uniform variates into a Markov chain. Let $g_1(\cdot)$ and $g_2(\cdot)$ be similar functions such that $g_1(U_{d_1}) = Z_{d_1}$ and $g_2(U_{d_2}) = Z_{d_2}$, with $g_1(\cdot)\neq g_2(\cdot)$ if $d_1 \neq d_2$. The relationship in \eqref{eq:spearman} and the assumption made in \eqref{eqn:gaussianpaircopula} together means that it is sufficient to estimate $\rho^*_{d_1d_2} = \rho^*(U_{d_1},U_{d_2})$ to obtain an estimate of $\rho_{d_1d_2} = \rho(W_{d_1},W_{d_2})$. If we denote the corresponding estimators as $\hat{\rho}^*_{d_1 d_2}$ and $\hat{\rho}_{d_1 d_2}$ respectively, the estimate $\hat{\rho}^*_{d_1d_2}$ can be obtained as the numerical solution to
\begin{align}\label{eqn:solveforrho}
  r_{d_1 d_2} &= \rho^*(g_1(U_{d_1}),g_2(U_{d_2});\rho^*_{d_1 d_2})\\
            &= \rho^*(Z_{d_1},Z_{d_2};\rho^*_{d_1d_2}),\nonumber
\end{align}
where $r_{d_1 d_2}$ is the sample Spearman correlation between states of the LHMM which is fixed given the data, and $\rho^*(Z_{d_1},Z_{d_2};\rho^*_{d_1d_2})$ is its population version. Note that it is not possible to invert the relationship in \eqref{eqn:solveforrho} and obtain an analytical expression for $\hat{\rho}^*_{d_1 d_2}$ as one perhaps would in a method of moments approach. However, given any value of $\rho^*_{d_1d_2}$ it is straightforward to generate data from the MMC and obtain a large sample estimate $r^*_{d_1 d_2}$ of $\rho^*(Z_{d_1},Z_{d_2};\rho^*_{d_1d_2})$. The Spearman correlation is not preserved by this transformation and $\rho^*(g_1(U_{d_1}),g_2(U_{d_2})) \neq \rho^*(U_{d_1},U_{d_2})$ except in trivial cases. \citet{MajumderThesis} has empirically shown that a monotonically increasing relationship exists between $\rho^*(g_1(U_{d_1}),g_2(U_{d_2})) \mbox{ and } \rho^*(U_{d_1},U_{d_2})$, and that $\rho^*(g_1(U_{d_1}),g_2(U_{d_2})) < \rho^*(U_{d_1},U_{d_2})$. The inequality is a consequence of $g_1(\cdot)$ and $g_2(\cdot)$ discretizing continuous variables $U_{d_1}$ and $U_{d_2}$ into $Z_{d_1}$ and $Z_{d_2}$ which are ordinal variables with 2 levels and possible ties. This attenuates the maximum and minimum values that the Spearman correlation between the 2 state processes can take. \citet{MhannaBauwens2012} have also demonstrated similar behaviour using empirical studies when Uniform variables are discretized to Bernoulli variables. The monotone relationship between $\rho^*(g_1(U_{d_1}),g_2(U_{d_2})) \mbox{ and } \rho^*(U_{d_1},U_{d_2})$ means that for a given target value of $r_{d_1d_2}$, it is possible to use a line search to identify the value of $\rho^*_{d_1d_2}$ which generates states with a sample Spearman correlation of $r^*_{d_1 d_2}$ arbitrarily close to $r_{d_1d_2}$. The $D(D-1)/2$ unique elements of $\Sigma$ can thus be estimated using pairs of state sequences. 

\subsection{Algorithm to estimate Gaussian copula parameters}\label{sec:copulalgo}
Recall that for the LHMM, The states for each sector's HMM are obtained using the Viterbi algorithm once the marginal parameters have been estimated. Let $\{r_{d_1 d_2}\}$ denote the observed Spearman correlations between the states of each $(d_1,d_2)\in \mathcal{D}^2$ pair of the $D$ component HMMs. This value is fixed given the marginal models and the data. Given the $n \times D$ matrix of states, the initial distribution $\myvec{\alpha}_{d}$, and the transition matrix $\mathbf{\Pi}_d$ for each $Z_d$, we want to construct a Gaussian copula that can generate an MMC with pairwise Spearman correlations $r^*_{d_1 d_2}$ coinciding with $\{r_{d_1 d_2}\}$. Let  $\hat{\rho}_{d_1d_2}$ be the estimate of the copula correlation in \eqref{eqn:gaussianpaircopula} between $(W_{d_1},W_{d_2})$, and let $\hat{\rho}_{d_1d_2}^*$ be the corresponding estimate of the Spearman correlation using \eqref{eq:spearman}. Since \eqref{eqn:solveforrho} cannot be rewritten as a function of $\rho^*_{d_1 d_2}$, we resort to a simulation approach to compute $\hat{\rho}_{d_1d_2}^*$ and $\hat{\rho}_{d_1d_2}$. We initialize $\hat{\rho}_{d_1 d_2}^*$ with $r_{d_1 d_2}$ for each pair of Markov chains $(Z_{d_1},Z_{d_2})$ and simulate an MMC from the Gaussian copula. We compute the pairwise Spearman correlations between the Markov chains in the MMC and denote them by $r_{d_1d_2}^*$. If $r_{d_1d_2}^*< r_{d_1d_2}$, we increment $\hat{\rho}_{d_1d_2}^*$ by a step size $\tau$ and repeat the process. We stop when $ \vert r_{d_1 d_2}^* - r_{d_1 d_2}\vert \leq \epsilon$, for some predefined tolerance $\epsilon$. The procedure is formalized in Algorithm 1 below.

\begin{algorithm}[H]
\SetAlgoLined
Segment $y_{1:K}$ into its $D$ sectors according to S\&P 500\\
Estimate marginal HMM parameters $\myvec{\alpha}_d$, $\mathbf{\Pi}_d$, and $\myvec{\theta}_d$ for sectors $d=1, \ldots, D$ using the Baum-Welch algorithm\\
Estimate $Z_{d,1}, \ldots Z_{d,n}$ using the Viterbi algorithm for sectors $d=1, \ldots, D$\\
Set step size $\tau$ and tolerance $\epsilon$\\
\For{sectors $(d_1,d_2) \ni d_1 = 1,\ldots,D-1$ and $d_2 = d_1+1,\ldots,D$ }{
Compute the observed Spearman correlation $r_{d_1 d_2}$ as in \eqref{eqn:solveforrho}\\
Initialize $\hat{\rho}_{d_1 d_2}^* = r_{d_1 d_2}$ \\
Initialize $r_{d_1 d_2}^* = 0$ \\
 \While{ $\vert r_{d_1 d_2}^* - r_{d_1 d_2}\vert  >\epsilon,$}{
 Increment $\hat{\rho}_{d_1 d_2}^*$ by $\tau$\\
 Compute Pearson correlation $\hat{\rho}_{d_1 d_2}$ from $\hat{\rho}_{d_1 d_2}^*$ using \eqref{eq:spearman}\\
 Generate correlated bivariate sequence from $N_2\biggl(\begin{pmatrix} 
 0\\
 0
 \end{pmatrix},\begin{pmatrix}
 1 & \hat{\rho}_{d_1 d_2}\\
 \hat{\rho}_{d_1 d_2} & 1
 \end{pmatrix}\biggr)$\\
 Use estimates of $\myvec{\alpha}_{d_2}$, $\myvec{\alpha}_{d_2}$, $\mathbf{\Pi}_{d_1}$, $\mathbf{\Pi}_{d_2}$, and the correlated sequences to generate synthetic states\\
 Calculate Spearman correlation $r_{d_1 d_2}^*$ of the synthetic states as an estimate of $\rho^*(Z_{d_1},Z_{d_2};\rho^*_{d_1d_2})$ as in \eqref{eqn:solveforrho}
 }
 }
 Construct correlation matrix $\hat{\Sigma}$ with off-diagonals $\hat{\rho}_{d_1 d_2}$ and diagonals set to 1\\
 \If {$\hat{\Sigma}$ is not positive definite}{
 Eigendecompose $\hat{\Sigma}$ as $\hat{\Sigma} = VRV^T$\\
 Replace negative and zero eigenvalues in $R$ with $10^{-6}$; call new matrix $R^*$\\
 Recalculate $\hat{\Sigma} = VR^*V^T$
 }
\caption{Algorithm to construct a Gaussian copula for an LHMM.}
\label{alg:statecopula}
\end{algorithm}
\noindent Since the entries of $\hat{\Sigma}$ are constructed independently, the resultant matrix is not guaranteed to be positive definite. The final steps of our algorithm ensures the positive-definiteness of $\hat{\Sigma}$. An alternative approach suggested by one of the anonymous referees is to add a similar small positive quantity to all diagonal elements of $\Sigma$. In cases when $\Sigma$ is high-dimensional and the eigendecomposition is computationally expensive, this would be a much faster way of ensuring the positive definiteness of $\Sigma$.

After $\Sigma$ has been estimated, we can use the correlation structure to re-estimate the marginal parameters $\myvec{\alpha}_d$, $\mathbf{\Pi}_d$, and $\myvec{\theta}_d$. One way of doing so is generating a sequence of states from the MMC and use the states as initial values in the Baum-Welch algorithm. Alternatively, we can re-estimate $\myvec{\alpha}_d$ and $\mathbf{\Pi}_d$ for all sectors from synthetic states generated from the MMC, and use the estimates as the initial distributions in the Baum-Welch algorithm to restimate all marginal parameters. For this study, we have followed the second approach.
\section{Stock Portfolio Selection using an LHMM}\label{sec:LHMM}
The return of the $k$\textsuperscript{th} stock over $n$ weeks is defined as follows,
\begin{equation}\label{R_stock}
R_k=\prod^n_{t=1} (1+Y_{k,t}).
\end{equation}
Our desired portfolio generates a high return with a low risk over a period of time, so we seek stocks with these characteristics as well.  To evaluate each of the stocks, we use the random variable $R_k$. 
Given a portfolio of $K$ stocks with allocations $\myvec{w} = (w_1,\ldots, w_K)$, its return over $n$ weeks is defined as,
\begin{equation}\label{R_o_P}
R(w_1,w_2,\ldots,w_K) = \sum^K_{k=1} w_k R_k,
\end{equation}
where the weight $w_k$ represents the proportion of the portfolio wealth invested in the $k$\textsuperscript{th} stock.
Thus, the expected return of a portfolio is given by
\begin{equation}\label{R_port}
E(R) = \sum^K_{k=1} w_k E(R_k).
\end{equation}
The variance of return is given by,
\begin{equation}\label{V_p_p}
V(R) = \sum^K_{k=1} w_k^2 \mathrm{Var}(R_k) + \sum^K_{k=1} \sum^K_{l \neq k}  w_k w_l \mathrm{Cov}(R_k, R_l)
\end{equation}
The goal of portfolio selection in this paper is to find the optimal allocation $\myvec{w}=(w_1,\ldots,w_k)$ with high reward $E(R)$ and relatively low risk $V(R)$ based on the results above. 
The optimal $\myvec{w}$ would maximize $E(R)$ while minimizing $V(R)$.  However, empirical evidence suggests that there exists a trade-off between $E(R)$ and $V(R)$ \cite[p.~200]{M19}. The most conservative approach would be to choose $\myvec{w}$ such that $V(R)$ is minimized, i.e.,
\begin{align}
   \myvec{w}_v = \arg \min_{\myvec{w}} V(R)  \text{, subject to } E(R) > 0 \text{ and } \sum^K_{k=1}w_k=1. 
\end{align}
Alternatively, \citet{M19} suggested that an optimal weight vector $\myvec{w}^*$ should maximizes $E(R)$ while $V(R)=v$,
\begin{equation*}
\myvec{w}^*(v) = \arg \max_{\myvec{w}} E(R)  \text{, subject to } V(R) = v \text{ and } \sum^K_{k=1}w_k=1.
\end{equation*}
The Lagrange multiplier method is used to find optimal allocations. The pairs of $E\{R\left(\myvec{w}^*(v)\right)\}$ and $v$ are referred to as the efficient ($R$, $V$) combinations by \citet{M52}. He claimed that a portfolio created based on an efficient combination is efficient, but did not suggest a specific combination to balance the reward and the risk.  \citet{JN19} suggested the following approach to find a vector of weights $\myvec{w}_b$ for a balanced portfolio,
\begin{equation*}
\myvec{w}_b = \arg \max_{\myvec{w}}  E(R)-q\sqrt{V(R)}  \text{, subject to } \sum^K_{k=1}w_k=1.
\end{equation*}
In this expression, $q$ functions as a tuning parameter which controls the trade-off between reward and risk.  A similar technique was also implemented in \citet{EH97}.  Assuming $R$ is approximately Normal, this technique maximizes the lower bound of a $95\%$ confidence interval of the return of a portfolio.  As $q$ gets higher in value, the resulting portfolio would accept less risk and prioritize more stable stocks.  As $q$ gets lower in value ($q \geq 0$), the portfolio would select stocks with higher return despite their higher volatility.  The choice of $q$ is based on an investor's willingness to take risk.  For the rest of the paper, we will assume $q=2$.

In practice, the stocks $Y_k$, for $k=1,\ldots, K$ will not be Normally distributed. We can transform them to Normal variates $Y^*_k$ using the Yeo-Johnson power transformation \citep{YeoJohnson}, and fit HMMs on the transformed variables $Y_k^*$. While analytical expressions analogous to $E(R)$ and $V(R)$ can be constructed based on $Y_k^*$, the quantities $E(R^*)$ and $V(R^*)$ do not have any meaningful interpretations. However, since we can generate data from the fitted LHMM, a simulation based approach allows us to recover data in the original scale.

Consider an LHMM fitted to the transformed stock returns $\mathbf{Y}^* = (Y_1^*, \ldots, Y_K^*)$ using the methodology described in Algorithm \ref{alg:statecopula}. We can simulate data from this model
- let us denote this simulated data by $\widehat{\mathbf{Y}}^* = (\widehat{Y}_1^*, \ldots, \widehat{Y}_K^*)$. Since $\widehat{Y}_k^*$ has the same distribution as $Y_k^*$, we use the inverse of the Yeo-Johnson transform to recover $\hat{Y}_k$ for $k=1,\ldots,K$. $\widehat{Y}_k$ are simulated stock price changes, and thus we can estimate $\hat{R}_k$ from this data using \eqref{R_stock}. If we simulate a large number of independent datasets from the fitted model, say $N$, we have for the $k$\textsuperscript{th} stock $N$ independent annual return samples $R_k^1, \ldots, R_k^N$. The vector of expectations $E(R)$ and the covariance matrix $V(R)$ can be computed from this data, and can be used for portfolio optimization.
\section{Building a Portfolio for 2016--17 from S\&P 500 Data}
\begin{table}
\small
\centering
\caption{S\&P 500 stocks per sector used to fit an LHMM using data from 2011-10-01 to 2016-09-30.}
\label{tab:sectors}
\begin{tabular}{lc}
\toprule
\textbf{Sector}        & \textbf{Number of Stocks}  \\\midrule
Communication Services & 8                          \\
Consumer Discretionary & 69                         \\
Consumer Staples       & 32                         \\
Energy                 & 28                         \\
Financials             & 67                         \\
Health Care            & 43                         \\
Industrials            & 69                         \\
Information Technology & 45                         \\
Materials              & 40                         \\
Real Estate            & 10                         \\
Telecommunications     & 7                          \\
Utilities              & 29                         \\
\textbf{Total}         & \textbf{447}              \\\bottomrule
\end{tabular}
\normalsize
\end{table}
We fit an LHMM to historical S\&P 500 data to create a portfolio for 2016--17 and evaluate its performance against the S\&P 500 index changes. 
As described in Section \ref{sec:LHMM}, the parameters of the LHMM are used to identify efficient $(R,V)$ combinations and use the associated weights to create a portfolio. Historical data for S\&P 500 stocks from 2011-10-01 to 2016-09-30 is used to build an LHMM with 12 Markov chains corresponding to the 12 sectors represented in the data. Stocks with records of fewer than 5 years are ignored; this leaves us with 447 stocks for the study, i.e., $K=447$. 

Table \ref{tab:sectors} shows the number of stocks available per sector that were used to fit the LHMM. The weekly stock price changes $\mathbf{Y}$ were made to undergo the Yeo-Johnson power transformation; the resulting variable $\mathbf{Y}^*$ is Normally distributed and thus meets the distributional assumptions for the LHMM. The HMMs were fitted using the packages \texttt{depmixS4} \citep{depmixS4} and \texttt{hmmr} \citep{hmmr} on \texttt{R 4.0.x}. For each sector, the B-W algorithm was restarted 20 times with random starting values. Parameter estimates from each of the 20 random restarts were compared on the basis of their Bayesian information criterion (BIC), with lower BIC values corresponding to higher likelihoods. The model which provided the lowest BIC values was chosen as the final model for each sector.

Once HMMs have been fitted to each sector's data, the most likely sequence of states was obtained using the Viterbi algorithm. The states were labeled such that State 1 is the bear state for each HMM and State 2 is the bull state, and the target Spearman correlation matrix was computed based on each pair of state processes. Next, a Gaussian copula which can generate synthetic states with the same Spearman correlation was constructed using Algorithm~\ref{alg:statecopula}. Synthetic state sequences from the copula as used to re-estimate $\myvec{\alpha}_d$, $\mathbf{\Pi}_d$, and $\myvec{\theta_d}$ for $d = 1,\ldots,D$; these estimates now take into account the correlation structure between $Z_1,\ldots,Z_D$.

Once all LHMM parameters have been estimated, 10000 datasets of 5 years ($n=260$ weeks) each were simulated from this fitted model, and the emissions $\hat{\mathbf{Y}}^*$ of the simulated data were transformed back to their original scale $\hat{\mathbf{Y}}$ using the inverse of the Yeo-Johnson transformation. The gains $R_{k}^i$ are computed for the $k=1,\ldots,447$ stocks for the $i=1,\ldots, 10000$ datasets. This gives us a $10000 \times 447$ matrix of $R_{k}^i$ values; $E(R)$ and $V(R)$ can be computed from this matrix. These simulated values of $E(R)$ and $V(R)$ were used for constrained optimization to obtain the optimum weight vector $\myvec{w}_v$ which minimizes $V(R)$ subject to $E(R)>0$, and $\myvec{w}_b$ which maximizes $E(R) - 2\sqrt{V(R)}$. We denote the latter as a balanced portfolio assignment since it balances the expected return with the uncertainty surrounding it, and portfolios for the period 2016-10-01 to 2017-09-30 can be built based on $\myvec{w}_b$ and $\myvec{w}_v$. We repeated this entire process 100 times, to get 100 estimates of the weights, $\myvec{w}_b^{(1)},\ldots,\myvec{w}_b^{(100)}$ and $\myvec{w}_v^{(1)},\ldots,\myvec{w}_v^{(100)}$. These are used to construct confidence intervals for the $\%$-age gains based on our method, and we evaluated their performance against the capital gains from 2016-10-01 to 2017-09-30.
\begin{table}
\small
\centering
\caption{Actual gains in \%-age in the one-year period from 2016-10-01 to 2017-09-30 based on four different portfolios. 95\% bootstrap confidence intervals are provided in parantheses. The corresponding S\&P 500 gains for this time period is 18\%.}
\label{tab:returns}
\begin{tabular}{lcccc}\toprule
\multirow{2}{*}{Sector} & \multicolumn{2}{c}{\% gain from HMMs} & \multicolumn{2}{c}{\% gains from LHMM} \\
 & Min V(R) & Balanced & Min V(R) & Balanced \\\midrule
Communication Services & \begin{tabular}[c]{@{}c@{}}- 0.10\\ (-0.17,-0.03)\end{tabular} & \begin{tabular}[c]{@{}c@{}}0\\ (0,0)\end{tabular} & \begin{tabular}[c]{@{}c@{}}0\\ (0,0)\end{tabular} & \begin{tabular}[c]{@{}c@{}}0\\ (0,0)\end{tabular} \\
Consumer Discretionary & \begin{tabular}[c]{@{}c@{}}1.52\\ (1.37,1.67)\end{tabular} & \begin{tabular}[c]{@{}c@{}}0.19\\ (0.14,0.23)\end{tabular} & \begin{tabular}[c]{@{}c@{}}1.05\\ (1.83,1.28)\end{tabular} & \begin{tabular}[c]{@{}c@{}}0.34\\ (0.27,0.48)\end{tabular} \\
Consumer Staples & \begin{tabular}[c]{@{}c@{}}0.38\\ (0.27,0.50)\end{tabular} & \begin{tabular}[c]{@{}c@{}}0.60\\ (0.21,0.97)\end{tabular} & \begin{tabular}[c]{@{}c@{}}0.07\\ (0,0.12)\end{tabular} & \begin{tabular}[c]{@{}c@{}}1.96\\ (1.56,2.33)\end{tabular} \\
Energy & \begin{tabular}[c]{@{}c@{}}0.58\\ (0.47,0.70)\end{tabular} & \begin{tabular}[c]{@{}c@{}}0\\ (0,0)\end{tabular} & \begin{tabular}[c]{@{}c@{}}0.70\\ (0.51,0.90)\end{tabular} & \begin{tabular}[c]{@{}c@{}}0\\ (0,0)\end{tabular} \\
Financials & \begin{tabular}[c]{@{}c@{}}0.70\\ (0.49,0.88)\end{tabular} & \begin{tabular}[c]{@{}c@{}}3.63\\ (3.25,4.08)\end{tabular} & \begin{tabular}[c]{@{}c@{}}0.27\\ (0.18,0.45)\end{tabular} & \begin{tabular}[c]{@{}c@{}}2.03\\ (1.55,2.56)\end{tabular} \\
Health Care & \begin{tabular}[c]{@{}c@{}}2.52\\ (2.26,2.73)\end{tabular} & \begin{tabular}[c]{@{}c@{}}-0.33\\ (-0.49,-0.13)\end{tabular} & \begin{tabular}[c]{@{}c@{}}4.48\\ (4.04,4.93)\end{tabular} & \begin{tabular}[c]{@{}c@{}}-0.99\\ (-1.27,-0.68)\end{tabular} \\
Industrials & \begin{tabular}[c]{@{}c@{}}0.33\\ (0.18,0.50)\end{tabular} & \begin{tabular}[c]{@{}c@{}}0.38\\ (0.34,0.44)\end{tabular} & \begin{tabular}[c]{@{}c@{}}0.43\\ (0.22,0.57)\end{tabular} & \begin{tabular}[c]{@{}c@{}}0.16\\ (0.09,0.23)\end{tabular} \\
Information Technology & \begin{tabular}[c]{@{}c@{}}0.69\\ (0.57,0.78)\end{tabular} & \begin{tabular}[c]{@{}c@{}}7.18\\ (6.59,7.87)\end{tabular} & \begin{tabular}[c]{@{}c@{}}0.35\\ (0.26,0.44)\end{tabular} & \begin{tabular}[c]{@{}c@{}}9.29\\ (8.54,9.95)\end{tabular} \\
Materials & \begin{tabular}[c]{@{}c@{}}1.21\\ (1.06,1.39)\end{tabular} & \begin{tabular}[c]{@{}c@{}}0\\ (0,0)\end{tabular} & \begin{tabular}[c]{@{}c@{}}0.37\\ (0.20,0.55)\end{tabular} & \begin{tabular}[c]{@{}c@{}}0\\ (0,0)\end{tabular} \\
Real Estate & \begin{tabular}[c]{@{}c@{}}0.07\\ (-0.03,0.17)\end{tabular} & \begin{tabular}[c]{@{}c@{}}0\\ (0,0)\end{tabular} & \begin{tabular}[c]{@{}c@{}}0.16\\ (-0.03,0.37)\end{tabular} & \begin{tabular}[c]{@{}c@{}}0\\ (0,0)\end{tabular} \\
Telecommunications & \begin{tabular}[c]{@{}c@{}}0.54\\ (0.45,0.65)\end{tabular} & \begin{tabular}[c]{@{}c@{}}0\\ (0,0)\end{tabular} & \begin{tabular}[c]{@{}c@{}}0.40\\ (0.20,0.49)\end{tabular} & \begin{tabular}[c]{@{}c@{}}0\\ (0,0)\end{tabular} \\
Utilities & \begin{tabular}[c]{@{}c@{}}0.51\\ (0.39,0.66)\end{tabular} & \begin{tabular}[c]{@{}c@{}}0.49\\ (0.41,0.58)\end{tabular} & \begin{tabular}[c]{@{}c@{}}-0.12\\ (-0.21,0)\end{tabular} & \begin{tabular}[c]{@{}c@{}}-0.06\\ (-0.09,-0.05)\end{tabular} \\
\textbf{Total} & \textbf{\begin{tabular}[c]{@{}c@{}}8.97\\ (8.58,9.31)\end{tabular}} & \textbf{\begin{tabular}[c]{@{}c@{}}12.14\\ (11.29,13.23)\end{tabular}} & \textbf{\begin{tabular}[c]{@{}c@{}}8.11\\ (7.57,8.67)\end{tabular}} & \textbf{\begin{tabular}[c]{@{}c@{}}12.72\\ (11.60,13.58)\\\end{tabular}}\\\bottomrule
\end{tabular}
\normalsize
\end{table}

A second model was also considered, where we had the 12 marginal HMMs but did not have the Gaussian copula to specify an LHMM. While we also wanted to consider a baseline model where all 447 stocks were modeled using a single state process, numerical issues prevented the model from converging consistently when using random restarts. Table \ref{tab:returns} shows the performance of the two portfolios each for the HMMs and the LHMM compared with the S\&P~500 capital gains. For each sector, the first row provides the mean \%-age gains during the one year test period, and the second row provides the corresponding 95\% bootstrap confidence interval. If our aim is to just minimize risk, the LHMM does not provide better returns compared to individual HMMs. This approach results in a diversified portfolio for both models, where nearly every sector contributes to the annual gains. On the other hand, trying to balance expected return and risk leads to portfolios concentrated around a few sectors. In particular, Information Technology stocks were the single largest contributer to the annual gains for both the HMMs and the LHMM in our study. The balanced portfolios have higher annual gains compared to the portfolio which minimizes the variance, and the one based on the LHMM has the highest gain among all portfolios constructed, with a mean of 12.72\% with a confidence interval of (11.60\%, 13.58\%). If our primary goal is to balance return and risk, the LHMM which better encapsulates market dynamics by allowing the different state processes to evolve jointly, provides better overall returns.
\begin{table}
\small
\centering
\caption{Mean and standard deviation (SD) of the number of transactions for each type of portfolio based on 100 independent estimates of portfolio weights.}
\label{tab:numtransactions}
\begin{tabular}{lcccc}\toprule
 & \multicolumn{2}{c}{\begin{tabular}[c]{@{}c@{}}Number of transactions\\ for HMM portfolios\end{tabular}} & \multicolumn{2}{c}{\begin{tabular}[c]{@{}c@{}}Number of transactions\\ for LHMM portfolios\end{tabular}} \\
\textbf{} & Min V(R) & Balanced & Min V(R) & Balanced \\\midrule
\textbf{Mean} & 134.45 & 37.47 & 48.84 & 33.43 \\
\textbf{SD} & 3.44 & 1.00 & 1.56 & 1.03\\\bottomrule
\end{tabular}
\normalsize
\end{table}

Since we are demonstrating portfolio construction for a single year (2016--2017), the number of non-zero weights in our allocations correspond to the number of transactions for the entire year. This is another important metric to consider when comparing algorithms for portfolifo construction. The 100 different sets of weights in our case study thus correspond to 100 estimates of the number of transactions for each of the 4 approaches to portfolio selection considered here. Table \ref{tab:numtransactions} lists the mean and standard deviation for the number of transactions. We note that the LHMM based portfolios require fewer transactions than corresponding portfolios constructed from independent HMMs. In particular, for the portfolio which minimizes risk, the LHMM portfolio requires fewer than half the number of transactions as the independent HMMs portfolio. If we are constrained by the number of allowed transactions, the LHMM portfolio is more likely to produce higher returns based on our empirical studies with S\&P 500 data.

\section{Discussion}
One of the key numerical challenges for fitting HMMs to large datasets using the B-W algorithm is that they often have trouble converging even under repeated random restarts. Using an LHMM allowed us to sidestep this issue to a large extent, since we went from trying to fit a 447-dimensional emission process to at most a 69-dimensional emission process. The LHMM also allows the market dynamics for each sector to evolve in a dependent manner without needing every stock to be in the same state at every time point. A similar form of heterogeneity can also be induced if we increase the number of states, but interpreting a larger number of states can be difficult. Increasing the number of states also increases the number of emission distribution parameters significantly. Extending to a multivariate state process, however, does not result in an increase in the number of emission distribution parameters and a relatively modest increase in the number of state process parameters. 

One of the assumptions that is made in this paper is that the stock price changes for different stocks within a sector are distributed as independent Normal variables given the state, as shown in \eqref{eqn:lhmmEmissions}. This rarely holds in practice, and something akin to a power transform is necessary to meet the assumption. However, even if the emission distribution of each stock's price changes is individually Normal, it still fails to adequately capture the correlation within the emission process. Ideally, we would want to model the emissions for each sector (either in its original scale of measurement or in a power-transformed scale so as to ensure Normality) as a multivariate Normal distribution, which would allow us to explicitly parameterize the correlation between the weekly gains for different stocks. We were actually able to do this for sectors with a small number of stocks, but faced computational issues for some of the larger sectors. It might be possible to estimate multivariate Normal parameters for the larger sectors if our data is extended to be longer than 260 weeks. However, the market dynamics do change over time and extending the length of the data might have other negative consequences. This is one aspect that we want to address in future work. In particular, a variational Bayes approach \citep{mcgroryetal} where we can assign priors could potentially alleviate many of the numerical issues associated with B-W parameter estimation.

\section*{Acknowledgements} 
The hardware used in the computational studies is part of the UMBC High Performance Computing Facility (HPCF). The facility is supported by the U.S. National Science Foundation through the MRI program (grant nos.~CNS--0821258, CNS--1228778, and OAC--1726023)
and the SCREMS program (grant no.~DMS--0821311), with additional substantial support from the University of Maryland, Baltimore County (UMBC). 
See \url{hpcf.umbc.edu}
for more information on HPCF and the projects using its resources.
Reetam Majumder was supported by the Joint Center for Earth Systems Technology and by the HPCF as a Research Assistant.
\bibliographystyle{rss}
\bibliography{sn-bibliography}


\end{document}